\documentstyle[aps,prb]{revtex}

	\def\Journal#1#2#3#4{{#1} {\bf #2}, #3 (#4)}

	\def\PRL{Phys. Rev. Lett.}
	\def\PRB{Phys. Rev. B}
	\def\ZPB{Z. Phys. B}

	\def\La{La$_2$CuO$_{4 + y}$ }
	\def\LaSr{La$_{2-x}$Sr$_x$CuO$_4$ }
	\def\LaPure{La$_2$CuO$_4$ }
	\def\LaOptimal{La$_{1.85}$Sr$_{0.15}$CuO$_4$ }

	\def\Lans{La$_2$CuO$_{4 + y}$}          	
	\def\LaSrns{La$_{2-x}$Sr$_x$CuO$_4$}    	
	\def\LaPurens{La$_2$CuO$_4$}            	
	\def\LaOptimalns{La$_{1.85}$Sr$_{0.15}$CuO$_4$}	

	\def\tc{$T_c$ }
	\def\tcns{$T_c$}                        
	\def\tm{$T_m$ }
	\def\tmns{$T_m$}                        
	\def\et{{\it et al. }}
	\def\etns{{\it et al.}}                 

\begin{document}
\input epsf.sty
\twocolumn[\hsize\textwidth\columnwidth\hsize\csname %
@twocolumnfalse\endcsname
\draft
\widetext


\title{Neutron-scattering study of spin-density wave order in the
superconducting state of excess-oxygen-doped La$_2$CuO$_{4+y}$}

\author{Y. S. Lee, R. J. Birgeneau, and M. A. Kastner}
\address{Department of Physics and Center for Materials Science and
Engineering, Massachusetts Institute of Technology,
Cambridge, Massachusetts 02139}

\author{Y. Endoh and S. Wakimoto}
\address{Department of Physics, Tohoku University, Aramaki Aoba, Sendai 980-77, Japan}

\author{K. Yamada}
\address{Institute for Chemical Research, Kyoto University, Gokasho, Uji 610-0011, Japan}

\author{R. W. Erwin and S.-H. Lee}
\address{National Institute of Standards and Technology, Gaithersburg, Maryland 20899}

\author{G. Shirane}
\address{Department of Physics, Brookhaven National Laboratory, Upton,
New York 11973}

\date{\today}
\maketitle

\vspace{-3.4in}
\hspace{5in}
Phys. Rev. B (01Aug99)

\vspace{3.3in}
\vspace{-1ex}

\begin{abstract}

We report neutron scattering measurements of spin density wave order
within the superconducting state of a single crystal of predominately
stage-4 \La with a \tcns(onset) of 42 K.  The low temperature elastic
magnetic scattering is incommensurate with the lattice and is
characterized by long-range order in the copper-oxide plane with the
spin direction identical to that in the insulator.  Between
neighboring planes, the spins exhibit short-range correlations with a
stacking arrangement reminiscent of that in the undoped
antiferromagnetic insulator.  The elastic magnetic peak intensity
appears at the same temperature within the errors as the
superconductivity, suggesting that the two phenomena are strongly
correlated.  These observations directly reveal the persistent
influence of the antiferromagnetic order as the doping level increases
from the insulator to the superconductor.  In addition, our results
confirm that spin density wave order for incommensurabilities near 1/8
is a robust feature of the \LaPurens-based superconductors.

\end{abstract}

\


\phantom{.}
]
\narrowtext

\section{Introduction}

In the copper-oxide superconductors, the multiple roles played by the
electrons continue to defy a comprehensive theoretical understanding.
As carriers of both spin and charge, the strongly interacting
electrons are responsible for both the unconventional microscopic
magnetic behavior and the unusual macroscopic electronic properties of
the doped cuprates.  Localized antiferromagnetism is a dominant
feature of the cuprate phase diagram at low doping levels, while
conventional itinerant-electron behavior appears to obtain in the
overdoped, non-superconducting regime.  The intermediate doping levels
are dominated by neither, and this is where high-temperature
superconductivity occurs.  Over the past decade, experimental probes
such as nuclear magnetic resonance (NMR), muon spin resonance
($\mu$SR), and neutron scattering have revealed a fascinating
interplay between the spin fluctuations and the superconductivity in
the lamellar copper-oxides.  In particular, neutron scattering
experiments have conclusively shown that dynamic antiferromagnetic
correlations are a robust feature of the cuprates in both the
superconducting and normal
states.~\cite{Kastner,Yamada-doping,Wells-science,Aeppli-singular,Fong}

Additionally, it is becoming increasingly apparent that incommensurate
spin fluctuations are universal to the high \tc superconductors.  The
low-energy magnetic excitations in La$_{2-x}$Sr$_x$CuO$_{4+y}$
superconductors have been extensively studied, and the observed spin
fluctuations are characterized by wavevectors which are incommensurate
with the lattice.~\cite{Yamada-doping,Wells-science} These modulated
spin fluctuations persist in both the normal and superconducting
states, with a suppression of their intensity occurring below \tcns.
Yamada and coworkers~\cite{Yamada-doping} have shown in \LaSr that the
incommensurability $\delta$ increases progressively with the doping
level for $x\geq0.05$ with the simple empirical relation $\delta\simeq
x$, but saturates near a value of $\delta\simeq1/8$ above a doping
level of $x\simeq1/8$.  Recently, incommensurate magnetic scattering
has been seen in superconducting YBa$_2$Cu$_3$O$_{6+x}$ with the
explicit incommensurabilities consistent with those in
\LaSrns.~\cite{Mook-incomm} There still exist several competing models
to explain the spatially modulated spin correlations, with plausible
explanations ranging from spin-flip electron scattering across the
Fermi surface~\cite{Incomm-FS-1,Incomm-FS-2,Incomm-FS-3} to
fluctuating stripes of antiferromagnetic spins in a microscopically
phase-separated doped Mott insulator~\cite{Emery,Nayak,Zaanen}.

Much recent interest has focused on the observation of static
incommensurate magnetic ordering which coexists with superconductivity
in certain \LaPurens-based systems.  In
La$_{1.6-x}$Nd$_{0.4}$Sr$_x$CuO$_4$, Tranquada and
coworkers~\cite{Tranquada-Nd} have found evidence for a static
ordering of charge and spin with an incommensurate modulation.  They
propose a picture of antiferromagnetically ordered stripe regions
separated by charged domain walls which act as magnetic antiphase
boundaries.~\cite{Emery,Tranquada-Nd} The three
La$_{1.6-x}$Nd$_{0.4}$Sr$_x$CuO$_4$ samples studied with $x=0.12$,
0.15, and 0.20 all exhibit elastic incommensurate magnetic peaks with
onset temperatures of $\simeq$ 50~K, 44~K, and 15~K, respectively.
For the $x=0.12$ sample there is also evidence for charge ordering
with a somewhat higher onset temperature.  The incommensurabilities of
the static magnetic order are consistent with the dynamic spin
fluctuation incommensurabilities measured by Yamada
\etns~\cite{Yamada-doping} in samples of \LaSrns, that is, without
Nd$^{3+}$ co-doping.  Subsequent work~\cite{Tranquada-glass} has shown
that in the $x=0.12$ sample, La$_{1.48}$Nd$_{0.4}$Sr$_{0.12}$CuO$_4$,
the correlation length reaches its maximum value of
$\sim$200~\AA~below $\sim$30~K.  The $x=0.12$, 0.15, 0.20 samples are
all superconducting with onset
\tcns's of $\simeq$ 4~K, 11~K, and 15~K, respectively; here each
\tc is reduced with respect to its value in the absence of Nd$^{3+}$
co-doping.  This reduction of the superconducting \tcns's in
La$_{1.6-x}$Nd$_{0.4}$Sr$_x$CuO$_4$ is believed to have the same
origin as the suppressed superconductivity in La$_{2-x}$Ba$_x$CuO$_4$
with $x$ near 0.125.\cite{one-eighth} Both systems are tetragonal
below $\sim$70~K, in contrast to \LaSr which retains the orthorhombic
structure down to the lowest temperatures measured.  Noting that the
La$_{1.6-x}$Nd$_{0.4}$Sr$_x$CuO$_4$ samples with the largest magnetic
order parameters also have the lowest superconducting \tcns's,
Tranquada \et posit that the static stripe ordering and the
superconductivity compete with each other.  In addition, they
speculate that the low temperature tetragonal (LTT) phase is essential
for the static magnetic order in the superconductors.  Indeed, $\mu$SR
experiments have detected the reemergence of magnetic ordering in
La$_{2-x}$Ba$_x$CuO$_4$ and La$_{1.6-x}$Nd$_{0.4}$Sr$_x$CuO$_4$
samples at doping levels for which the superconductivity is suppressed
and the low temperature structure is
tetragonal.~\cite{Luke,Nachumi-usr}

Experiments on \LaSr show, however, that the existence of the LTT
structural phase is not a necessary condition for the appearance of
static magnetic order.  Based on their NMR studies of
\LaSr ($x\simeq0.115$) and La$_{2-x}$Ba$_x$CuO$_4$
($x\simeq0.125$), Goto \etns~\cite{Goto} conclude that magnetic order
exists even in the absence of the LTT structure.  More direct evidence
has come from recent neutron scattering experiments on \LaSr single
crystals with $x=0.10, 0.12, 0.13$, all of which have the low
temperature orthorhombic (LTO)
structure.~\cite{Suzuki,Kimura,Kimura-un} These measurements have
confirmed that incommensurate spin density wave (SDW) order exists at
doping levels near $x=1/8$ even in the absence of Nd$^{3+}$ co-doping.
In particular, the work of Kimura
\etns~\cite{Kimura} shows that the most dramatic effects occur for the
$x=0.12$ sample, for which the elastic peak width in their measurement
is resolution-limited, indicating a static magnetic correlation length
exceeding 200~\AA.  Surprisingly, \tcns(onset) and the SDW ordering
temperature, \tmns, coincide at $\sim$31~K within the errors, in
contrast to La$_{1.48}$Nd$_{0.4}$Sr$_{0.12}$CuO$_4$, in which \tm and
\tc are well separated.  It should be noted that, similar to
the previously mentioned compounds, there is evidence that the
superconductivity is suppressed somewhat in \LaSr for $x$ near
0.115,~\cite{Goto,dip-1,dip-2} though not to the same extent as in
La$_{2-x}$Ba$_x$CuO$_4$ for $x~\simeq~0.125$.

In view of the aformentioned studies, a fundamental issue needing
further clarification is the universality of the relationship between
the superconductivity, including any possible suppression of \tcns,
and the SDW ordering.  Although it is now clear that the LTT phase is
not essential for the establishment of the SDW, more experiments are
necessary to search for magnetic ordering in superconducting samples
which are orthorhombic rather than tetragonal.  Additionally, further
details of the spin structure of the magnetic order are essential to
elucidate the basic physics of these systems and to test the specifics
of competing models.  Most importantly, such studies may shed light on
the fundamental question of the appropriateness of an itinerant versus
doped Mott insulator model for the superconducting lamellar copper
oxides.  To address these issues, we have investigated a
superconducting single crystal of excess-oxygen-doped
\Lans.  For the \LaPurens-based superconductors, those doped with oxygen
achieve the highest superconducting \tcns's (up to 45~K) and are
believed to possess a relatively small degree of dopant
disorder.~\cite{Wells-science} The dopant oxygen ions are mobile down
to temperatures near $\sim$200~K.  Therefore, the oxygen interstitials
may find their equilibrium configuration and are annealed with respect
to the other major energetics of the system, at least at 200~K; here,
the magnetic exchange energy is $J~\simeq$~1500~K and the electronic
bandwidth is $W~\simeq~2J$.

In this paper, we present elastic and inelastic neutron scattering
data on a single crystal of \La with a superconducting transition
temperature onset of 42~K.  As we shall show, this crystal exhibits a
transition to incommensurate magnetic long-range order at a
temperature \tm coincident within the uncertainties with
\tcns(onset)~$\sim$~42~K.  We discuss in some detail the associated magnetic
structure.  The format of this paper is as follows: Section 2 contains
preliminary details about our measurements and about the
characterization of our \La sample.  The staging behavior is discussed
in Section 3. The results of the inelastic magnetic neutron scattering
experiments are described in Section 4.  Section 5 reports elastic
neutron scattering studies of the spin density wave ordering.
Finally, Section 6 contains a discussion of the results, especially in
the context of previous work.

\section{Preliminary Details}

In contrast to NMR and $\mu$SR techniques, which are local probes,
neutron scattering experiments uniquely measure the collective
behavior of spins or nuclei in a crystal.  The magnetic scattering
cross section of localized spins for unpolarized neutrons is directly
related to the Fourier transform of the space- and time-dependent
spin-pair correlation function.  Defining the neutron energy and
momentum transfers to the spin system as $\omega = E_f - E_i$ and
${\rm {\bf Q} = {\bf k_f} - {\bf k_i}}$, respectively, (here, we use
units in which $\hbar=k_B=1$) the inelastic neutron cross section is
given by
\begin{equation}
\frac{d^2\sigma}{d\Omega~dE} ~\sim~ f^2({\rm \bf
Q})~\frac{k_f}{k_i}~\sum_{\alpha,\beta}~(\delta_{\alpha \beta} -
\hat{Q}_\alpha \hat{Q}_\beta )~S^{\alpha \beta}({\rm \bf Q},
\omega),
\end{equation}
where $\alpha,\beta=x,y,z$, $f({\rm \bf Q})$ is the magnetic form
factor, and the dynamic structure factor is defined as
\begin{equation}
S^{\alpha \beta}({\rm \bf Q}, \omega) = \frac{1}{2\pi}\sum_{\bf
r}\int_{0}^{\infty}dt~e^{(i{\rm \bf Q} \cdot {\bf r} - \omega
t)}~\langle S^\alpha(0,0)S^\beta({\bf r}, t)\rangle.
\end{equation}
In ordered structures, for which the thermodynamic average of the spin
density appears at each lattice site, some portion of the scattering
occurs in elastic Bragg peaks.  The magnetic cross section for this
coherent elastic scattering is
\begin{eqnarray}
\left. \frac{d^2\sigma}{d\Omega dE} \right|_{Bragg} & \sim~~ &
f^2({\rm \bf Q}) \sum_{\alpha,\beta}~
(\delta_{\alpha \beta} - \hat{Q}_\alpha \hat{Q}_\beta ) \nonumber \\
& & \times \sum_{\bf d} e^{i{\rm \bf Q} \cdot {\bf d}}~\langle
S^\alpha(0)\rangle \langle S^\beta({\bf d})\rangle~\delta(\omega)
\end{eqnarray}
with ${\bf Q}={\bf G}$, a magnetic reciprocal lattice vector, and
$\sum_{\bf d}$ is a sum over the ions in the magnetic unit cell.  The
geometrical factor $(\delta_{\alpha \beta} -
\hat{Q}_\alpha \hat{Q}_\beta )$ in both cross sections is a consequence of the
dipolar interaction of the neutron with the spin and allows one to
determine the spin direction of static and fluctuating spins by
systematically varying the momentum transfer {\bf Q}.  The imaginary
part of the generalized susceptibility is related to the measured
dynamic structure factor via the fluctuation-dissipation theorem,
\begin{equation}
\chi^{''}({\bf Q},~\omega)~\frac{1}{1 - e^{-\omega/T}}~=~S({\bf Q},~\omega).
\end{equation}

In the case of an itinerant electron system, the measured structure
factor is related to the particle-hole excitations,
\begin{eqnarray}
\chi^{''}({\bf Q},~\omega) & ~=~ \sum_{\bf k} & (\langle n_{\bf
k\downarrow} \rangle - \langle n_{\bf k+Q\uparrow} \rangle) \nonumber \\
& & \times~\delta[E({\bf k+Q})-E({\bf k})-\omega].
\end{eqnarray}

The neutron scattering experiments described in this paper were
performed on the SPINS and BT9 triple-axis spectrometers located at
the NIST Center for Neutron Research in Gaithersburg, MD, and also on
the TOPAN triple-axis spectrometer located at the JAERI JRR-3M reactor
in Tokai, Japan.  Pyrolytic graphite (PG) crystals were used to
monochromate and analyze the neutron energies.  For the measurements
on the SPINS spectrometer, cold neutrons with an incident energy of
5~meV were selected, and a liquid-nitrogen cooled beryllium filter was
placed in the beam path to remove contamination from higher order
neutron energies.  For the measurements on the BT9 and TOPAN
spectrometers, thermal neutrons with incident energies of 13.7~meV or
14.7~meV were selected, and a PG filter was placed in the incident
beam to remove higher order neutron energies.  The sample was sealed
in an aluminum container in a He gas environment for thermal exchange.
A pumped $^4$He cryostat was used to control temperature from 1.4~K to
300~K.

We have, over the past few years, developed a reproducible method for
preparing single crystals of \La that are quite large, of high
purity, and with homogenous oxygen doping.  First, a crystal of
\LaPure is grown by the traveling solvent floating-zone method; the
crystal studied here has a volume of about 0.6 cm$^3$ with a crystal
mosaic of $0.2^\circ$ half-width at half-maximum.  In order to
incorporate a large quantity of excess oxygen into the crystal, we
employ an electrochemical doping technique.  Previous studies on
ceramic and small single crystal samples of \La have shown that
electrochemical
oxidation~\cite{Grenier,Chou-powder,Chou-crystal,Blakeslee} can
produce higher values of $y$ than methods involving annealing in high
oxygen pressure~\cite{pressure}. Recently, we have refined the
technique so that large single crystals can be homogeneously doped,
and thus are suitable for magnetic neutron scattering measurements,
which require large scattering volumes.  Wells
\etns~\cite{Wells-science} have investigated the phase diagram of \La
crystals and have reported neutron scattering measurements of the
low-energy magnetic excitations in superconducting
\Lans.  The superconducting crystals examined by these authors have \tcns's
of 31-32~K and represent the phase on the oxygen-rich boundary of the
first miscibility gap with an oxygen content of $y\simeq0.055$.

For the present studies we have successfully oxygenated a large single
crystal with an oxygen concentration well beyond the first miscibility
gap.  The bulk magnetic susceptibility of the \La crystal has been
measured with a SQUID magnetometer after cooling in zero field at a
magnetic field of 10~G.  As shown in Figure~1(A), the transition is
sharp with a \tcns(onset)~$\simeq~$42~K, significantly higher than the
maximum \tc of $\sim38$~K for \LaOptimalns.  The open symbols
represent measurements taken on the same sample used in the neutron
scattering experiments, both immediately after doping and \linebreak

\begin{figure}
\vskip -5mm
\centerline{\epsfxsize=3.5in\epsfbox{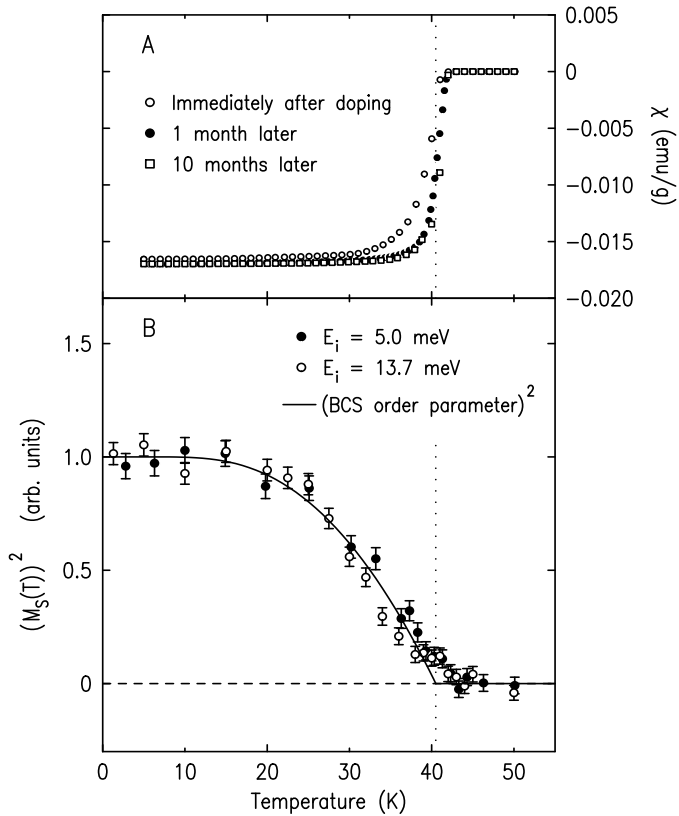}}
\vskip 5mm
\caption{
A) Bulk magnetic susceptibility measured after several time intervals
in an applied field of 10~G after cooling in zero field.  The open
symbols represent data taken with the same crystal used in the neutron
scattering experiments, while the closed symbols are taken on a small
piece which was broken off of the original crystal immediately after
doping.  The vertical dotted line denotes the approximate midpoint
temperature, \tcns(midpoint)$=40.5$~K, of the flux exclusion curve
after the one month anneal. B) Temperature dependence of the peak
intensity of the incommensurate elastic scattering, which is
proportional to the square of the magnetic order parameter $M_s$.  The
measurements have been performed with two different neutron energies
of 5.0~meV and 13.7~meV, yielding energy resolutions of 0.075~meV and
0.45~meV (half-width at half-maximum), respectively with our
particular choice of spectrometer collimations.  The solid line
denotes the square of the BCS order parameter, with $T_m=40.5~{\rm
K}=T_c$(midpoint).
}
\label{Figure-orderpar}
\end{figure}

\noindent 10 months later.  The solid symbols represent data taken on
a small piece broken off of the larger neutron scattering crystal.
The two pieces had identical temperature dependences for their
susceptibilities immediately after doping (not shown).  To correct for
their drastically different sample sizes, the susceptibilities of the
two samples have been normalized to coincide at the lowest
temperature.  Previous studies of ceramic samples of electrochemically
doped \Lans~\cite{Feng} have suggested that superconducting phases
with \tc~$>$~34~K may be unstable, exhibiting a significant diminution
of \tc after annealing at room temperature for several days.  We find
that for our single crystal sample, after a month-long anneal at room
temperature, the superconducting \tc does not decrease, but rather the
transition sharpens to a slightly higher temperature.  A 10 month
anneal produces only a minimal additional effect.  All of the neutron
scattering data shown in this paper have been taken after the one
month anneal.  It is apparent, therefore, that the superconducting
phase with a 42~K transition temperature onset in \La is a stable
phase.  Attempts to measure the Meissner effect by cooling in a
magnetic field yield a paramagnetic signal below \tcns, which is not
uncommon in studies of single crystal samples of \LaPurens-based
superconductors.~\cite{Tranquada-Nd,Chou-crystal}.  In accordance with
such previous studies, we believe that the zero-field-cooled
measurements provide reliable evidence for bulk
superconductivity.~\cite{Tranquada-Nd,Nachumi-usr,Ostenson-mag,Hirayama}

To estimate the oxygen doping level, thermogravitimetric analysis has
been performed on similarly doped single crystals from the same boule;
a separate sample was used so as not to de-oxygenate the sample used
in the neutron scattering measurements.  The oxygen weight loss yields
a doping level of $y~=~0.12~\pm0.005$, which is substantially higher
than the oxygen doping range for macroscopic phase separation into
superconducting and insulating phases.  Reported titration
measurements~\cite{Chou-powder,Li-carrier} on powder samples of \La
imply that an oxygen level of $y~\simeq~0.12$ corresponds to a hole
concentration of $p~\simeq~0.16$.  Indeed, SQUID magnetometer
measurements of the uniform magnetization at $T>T_c$ of our single
crystal sample reveal identical scaling behavior as seen in Sr$^{2+}$
doped samples~\cite{Johnston-chi-scaling,Nakano-chi-scaling}, with a
Sr$^{2+}$ doping level of $x\simeq0.14\pm0.02$.  Also, from $^{63}$Cu
nuclear quadrupole resonance (NQR) measurements, the NQR frequency as
well as the quantitative behavior of 1/T$_1$ match those found in
samples of \LaSr with $x\simeq0.15\pm0.02$.~\cite{Imai} Therefore, the
experimental evidence suggests that our crystal is a bulk
superconductor with a distribution of holes similar in density and
homogeneity to that of optimally doped, that is, $x\simeq0.15$, \LaSr
crystals.  The data shown in Figure~1(B) will be described later in
the discussion of the spin density wave peaks.

\section{Staging Behavior}

Our oxygen doped crystal is orthorhombic at low temperatures.  Undoped
\LaPure is also orthorhombic at low temperatures with the {\it Bmab} space
group symmetry, which differs from tetragonal symmetry because of the
tilt pattern of the CuO$_6$ octahedra.  One CuO$_6$ octahedron is
depicted in the unit cell diagram in Figure 2(A); the tilt pattern
causes a slight buckling of the CuO$_2$ plane as shown in Figure 2(B),
yielding the orthorhombic symmetry.  In \Lans, the interstitial
oxygens modify the structure along the $c$-direction by periodically
introducing planes across which the tilt pattern reverses.  The
segregation of excess oxygen into periodic planes, known as staging,
has been observed and explained previously by Wells
\etns~\cite{Wells-science,Wells-staging} for single crystals of \Lans.
Stage $n$ refers to a structure in which the tilt reversal occurs
periodically every $n$ CuO$_2$ layers.  Therefore, increasing the
excess oxygen content leads to structural phases with successively
smaller staging numbers $n$.  At low oxygen concentrations ($0.012~
{\Large \stackrel{_<}{_\sim}}~\delta~{\Large
\stackrel{_<}{_\sim}}~0.055$), for which phase separation into
oxygen-rich and oxygen-poor domains occurs, the oxygen-rich phase is
stage-6 with a superconducting \tc of
31-32~K~\cite{Wells-science,Blakeslee,Radaelli} while the oxygen-poor
phase is an insulating antiferromagnet.  For higher oxygen
concentrations ($\delta>0.055$), the crystal remains entirely oxygen
rich and stages 4, 3, and 2 are seen.~\cite{Wells-staging} Throughout
this paper, we use reciprocal lattice units (r.l.u.) for the {\it
Bmab} low temperature orthorhombic crystal structure, for which the
axes are depicted in Figure 2.  The reciprocal space ($H,K,L$)
components are given in units of ($2\pi/a,2\pi/b,2\pi/c$), where, at
low temperature, the in-plane lattice constants are
$a$~=~5.333~\AA~and $b$~=~5.401~\AA, and the out-of-plane lattice
constant is $c$~=~13.18~\AA.

We find, via neutron diffraction, that our \La crystal predominately
possesses the tilt structure corresponding to stage-4; however smaller
peaks corresponding to stage-2 and other staged phases are seen as
well.  In Figure 3, we show a scan through the staging superlattice
peaks which straddle the (014) {\it Bmab} peak position along the $L$
direction.  For each stage $n$ phase present, corresponding staging
peaks occur in pairs, split by $\pm 1/n$ from the {\it Bmab} position.
The line through the data represents the results of fits to pairs of
staging peaks using Gaussian lineshapes, convolved with the
resolution.  The data indicate that the components with stage-4,
fitted with $n=3.97(3)$, and stage-2, fitted with $n=2.01(1)$, account
for 80\% of the diffraction intensity in this scan.  Additionally, the
integrated intensity of the stage-4 peaks is 2.0 times larger than
that of the stage-2 peaks.  We find that the stage-2 peaks are
resolution limited in the $L$ direction, whereas the stage-4 peaks
have a non-zero width indicating a domain size of $\sim$240~\AA~along
the $L$ direction.  Both the stage-4 and stage-2 peaks are resolution
limited along the in-plane $K$ direction.  This comparison of the
integrated intensities of the staging peaks gives only an approximate
measure of the phase fractions present in the crystal.  To be more
accurate, one needs to account for the structure factor of the staged
phases, which depends significantly on the amplitude of the tilting of
the CuO$_6$ octahedra.  However, to a first approximation, assuming a uniform
amplitude of the octahedral tilt throughout the crystal, the
magnitudes of the structure factors for the staged phases are nearly
equal.  Minority phases with stage-6, stage-3, and undoped {\it Bmab}
structures account for the rest of the diffraction peaks.  This
confirms that integer values for staging numbers appear to give the
most stable phases, and that stage-5, which has yet to be observed in
high quality single crystals, is probably not stable.

Doping with mobile oxygen interstitials has the advantage of yielding
a lower degree of dopant \linebreak

\begin{figure}
\vskip -5mm
\centerline{\epsfxsize=3.5in\epsfbox{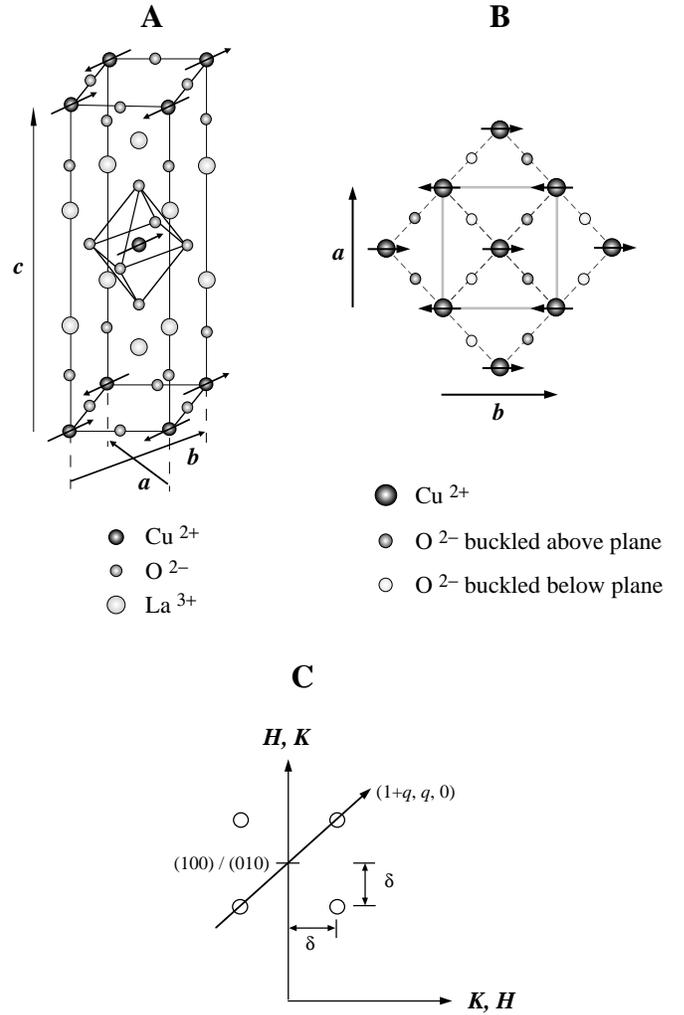}}
\vskip 5mm
\caption{ A) Unit cell of \LaPure depicting the orthorhombic axes.  B)
Diagram of the copper-oxygen plane, highlighting the 2D orthorhombic
unit cell.  C) Reciprocal space map of the positions of the {\it
inelastic} incommensurate magnetic peaks in superconducting
La$_{2-x}$Sr$_x$CuO$_{4+y}$.  The trajectory of the inelastic
scattering scans in Figure 4A is denoted by the line ($1+q$,~$q$, 0).}
\label{Figure-structural}
\end{figure}

\noindent disorder.  In fact, we have recently succeeded in observing
scattering directly from the ordered distribution of interstitial
oxygen, at reciprocal lattice positions distinct from the staging
superlattice peaks.~\cite{Lee-Tewary} The annealed character of the
oxygen disorder is revealed in measurements linking a reduction of the
superconducting \tc to the loss of the ordered oxygen
lattice.~\cite{Lee-Tewary} A possible disadvantage of having mobile
oxygens is that it may be easier for the crystal to become
macroscopically inhomogenous by phase separating into regions with
different hole concentrations.  The presence of multiple staged phases
in our crystal implies that the oxygen concentration may vary between
the macroscopic domains with different staging numbers.  It has been
shown that the oxygen \linebreak

\begin{figure}
\vskip -5mm
\centerline{\epsfxsize=3.0in\epsfbox{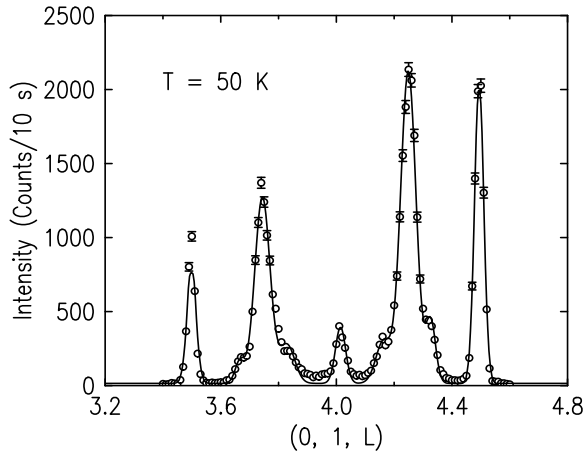}}
\vskip 5mm
\caption{
Scan along the $L$ direction through the staging superlattice
positions at $T=50$~K measured with 5~meV neutrons and horizontal
collimations of \mbox{$30^\prime-40^\prime-S-20^\prime-B$}, where $S$
denotes the sample and $B$ is a blank collimator.  The solid line
denotes the results of a fit to pairs of staging peaks centered about
the (014) position in addition to a small $Bmab$ peak at (014),
convolved with the instrumental resolution.  }
\label{Figure-staging}
\end{figure}

\noindent
concentration sharply increases between the stage-6 phase, with
$y=0.055$, and the stage-4 phase, with $y\simeq0.11$.~\cite{Blakeslee}
However, it is not known how the interstitial oxygen concentration
varies between the stage-4, 3, and 2 phases, which account for most of
the volume of our crystal.  Other electrochemical studies on ceramics
and single crystals of \La report that the oxygen doping level
produced by this method saturates at a maximum value of $y \sim
0.12$.~\cite{Chou-powder,Chou-crystal,Li-carrier} Therefore, it
appears likely that the oxygen concentration does not vary
substantially between the stage-4, 3, and 2 phases.  This is
consistent with the previous measurements of Wells
\etns~\cite{Wells-staging} which show that the $c$-axis lattice
constant is the same for a crystal with a single stage-4 phase and one
with simultaneous stage-4, 3, and 2 phases.  High resolution
measurements of the $c$-axis lattice constant for our crystal give no
indication that the sample has more than a single oxygen doped phase.
Thus, we conclude that our sample is both highly oxygen doped and
completely oxygen-rich throughout its bulk.  Also, we reiterate that
measurements which depend on hole density indicate that our \La sample
behaves similarly to \LaSr with $x\sim0.15$.  Therefore, the doped
holes in this oxygen-rich sample seem to be as uniformly distributed
throughout the bulk as those in Sr-doped \LaSrns, and have an
effective hole concentration of $n_h
\simeq 0.15\pm0.02$.  This is consistent with our observation of a
single, sharp superconducting transition at \tcns(onset)~=~42~K.  In
addition, as will be discussed next, our neutron scattering
meausurements of the magnetism are consistent with scattering from a
sample with a single hole concentration.

We should note that we have recently developed a technique for
electrochemically doping \La which appears to produce pure stage-4
material.~\cite{Lee-Tewary} However, at the present time this
uniformity in the staging number is offset by a slight decrease in \tc
and increased oxygen disorder.  We plan neutron scattering experiments
on these pure stage-4 systems in the near future.

\section{Inelastic Magnetic Neutron Scattering}

In previous inelastic neutron scattering work on stage-6
\La~\cite{Wells-science}, incommensurate magnetic excitations have been
confirmed to be universal in the \LaPurens-based superconductors.  The
magnitude and direction of the incommensurate wavevector follows the
same pattern as that for superconducting Sr$^{2+}$ doped
\LaPurens.~\cite{Yamada-doping,Wells-science} In particular, the behavior of
$\chi^{''}({\bf Q}, \omega)$ for \La with $T_c=31$~K is essentially
identical to that for underdoped \LaSr with similar \tcns, in which a
partial suppression of the intensity of the spin excitations is seen
as the temperature is reduced below the superconducting \tc.  This
contrasts with the behavior in optimally doped \LaSr with $x=0.15$, in
which a complete gap is observed in the magnetic excitation spectrum
in the low temperature superconducting state for energies
$\leq~3.5$~meV.~\cite{Yamada-gap} This difference is probably related
to the apparent absence of short-range or long-range SDW order in
La$_{1.85}$Sr$_{0.15}$CuO$_4$~\cite{Kimura}, whereas in the underdoped
samples SDW order occurs in the superconducting state.

We have performed inelastic neutron scattering measurements on our \La
sample probing the magnetic excitations with an energy transfer of
2~meV both above and below the superconducting transition.  The
low-energy spin fluctuations in \LaPurens-based superconductors
consist of four peaks equally displaced from the antiferromagnetic
zone center by an incommensurate wavevector $\delta$ along each of the
four Cu-O-Cu directions.  This arrangement is depicted in Figure~2(C),
where the $(100)$ and $(010)$ orthorhombic positions would be
equivalent to each other if the crystal were tetragonal, and both
correspond to the ($\frac{1}{2}$$\frac{1}{2}$0) position in tetragonal
notation.  Scans through two of the incommensurate peak positions have
been taken along the trajectory (1+$q$,~$q$,~0) as shown in
Figure~2(C).  Representative data collected at $T=2$~K and $T=44$~K
are plotted in Figure 4(A).  The solid lines are the results of fits
to 2D Lorentzians convolved with the instrumental resolution; the
Lorentzian width is found to be less than 0.008~\AA$^{-1}$ half-width
at half-maximum (HWHM) and is independent of temperature within the
errors.  Such a narrow width is also seen for \LaSr crystals with a
doping level close to 1/8.~\cite{Yamada-doping} (Note that there is a
weak contaminant peak at $q\simeq0.15$.)  The incommensurability is
found to be $\delta~\simeq~0.121(2)$.  The temperature dependence of
$\chi^{''}({\bf Q}_\delta, \omega)$ which is presented in Figure~4B,
is strikingly different from that for the optimal Sr$^{2+}$ doped
material \LaOptimal \linebreak

\begin{figure}
\vskip -5mm
\centerline{\epsfxsize=3.0in\epsfbox{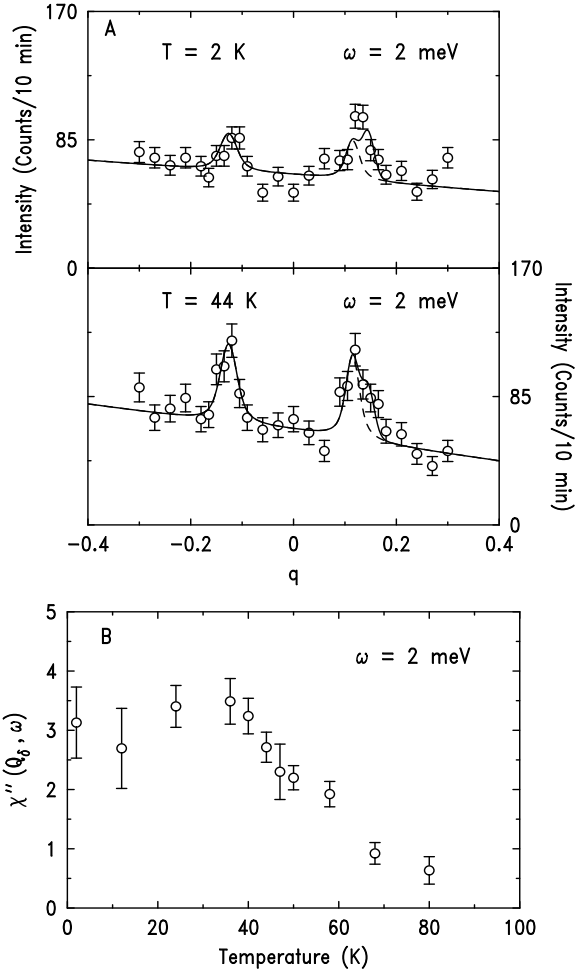}}
\vskip 5mm
\caption{
A) Inelastic scans through the incommensurate positions along the
trajectory given in Figure 2C at $T=2$~K and $T=44$~K for an energy
transfer of $\omega~=~$2~meV, measured with 14.7~meV neutrons and
horizontal collimations of \mbox{$B-60^\prime-S-60^\prime-B$}.  The
solid lines denote the results of fits to the cross section described
by four incommensurate peaks, together with a sloping background and a
weak, temperature-independent, contaminant peak on the shoulder of the
high-$q$ peak.  The dashed line at the high-$q$ position denotes the
magnetic signal.  B) The imaginary part of the susceptibility,
$\chi^{''}({\bf Q}_\delta,
\omega)$, extracted from the intensity of the scattering at the
incommensurate positions via Eq. 4.
}
\label{Figure-inelastic}
\end{figure}

\noindent
with \tcns~=~37.5~K.~\cite{Yamada-gap} We see
that, above the superconducting transition temperature, the measured
susceptibility increases continuously as the temperature decreases.
However, below \tc the dynamic susceptibility levels off and
approaches a non-zero constant value as $T\rightarrow0$, whereas for
\LaOptimal the intensity diminishes to zero at low temperature.  This
is a clear difference between oxygen doping and Sr$^{2+}$ doping; even
though both \tcns 's appear to be optimal, the low energy magnetic
excitations behave differently in the superconducting state.  As we
shall discuss below, the non-zero intensity at low temperatures in the
\La sample arises from scattering from spin wave excitations
associated with the SDW.  A few remarks are appropriate at this point.
First, the 2~meV inelastic peaks are sharp in reciprocal space, and
only a single set is seen.  Previous systematic studies in \LaSr show
that the incommensurability of the inelastic scattering depends
sensitively on the hole doping level for $x
\stackrel{_<}{_\sim}$~0.125 and saturates around 1/8 for doping levels
$x \stackrel{_>}{_\sim}$~0.125.  Therefore, since the observed value
of the inelastic incommensurability $\delta$ in our sample is near
1/8, this implies that the hole doping level is also near 1/8 or
possibly higher, which would be consistent within the errors with our
deduction above that $n_h=0.15\pm0.02$.  Additionally, since the width
we measure is as small as that of the narrowest peaks observed in
\LaSrns, the incommensurate inelastic scattering is strong evidence
for a homogenous distribution of holes in our sample.  Even if there
were a distribution of hole concentrations in our crystal, the
observed magnetic scattering would be attributable to at least one of
these concentrations, and all such high oxygen doping concentrations
are known to be superconducting.~\cite{Grenier,Chou-powder} Therefore,
we conclude that these sharp low-energy magnetic fluctuations and the
superconductivity coexist in the same doped phase.

\section{Spin Density Wave Ordering}

The most important finding of this study is that we observe sharp
elastic magnetic scattering at low temperatures.  The observed elastic
peak positions are shown in Figure~5.  Due to the twinning of this
orthorhombic crystal, which is tetragonal at high temperatures, the
neutrons simultaneously scatter from four twin domains.  Two of the
twin domains have different in-plane lattice constants along the same
direction; thus we simultaneously see both the $(020)$ and $(200)$
nuclear Bragg peaks in a single longitudinal scan.  In addition, each
of these domains has another twin counterpart which is rotated about
the $c$-axis by a small angle ($\sim0.7^\circ$) which is proportional
to the size of the orthorhombic distortion.  Since the twinning can be
precisely characterized via scattering from the nuclear Bragg peaks,
there is little ambiguity in interpreting the magnetic scattering.

Almost all previous neutron scattering measurements of the {\it
inelastic} incommensurate peaks have used a relatively wide resolution
to optimize the signal to background ratio, since the intrinsic signal
is found to be extended in energy.  In our measurements of the {\it
elastic} scattering, we find that the signal is narrow in both energy
and wavevector.  Therefore, we employ a relatively narrow resolution
to obtain a precise measurement of the peak positions and widths in
reciprocal space.  Taking a narrow mesh of scans in the ($HK0$)
scattering plane, we find two twin-related sets of four incommensurate
peaks which share a common center at the \linebreak

\begin{figure}
\vskip -10mm
\centerline{\epsfxsize=3.5in\epsfbox{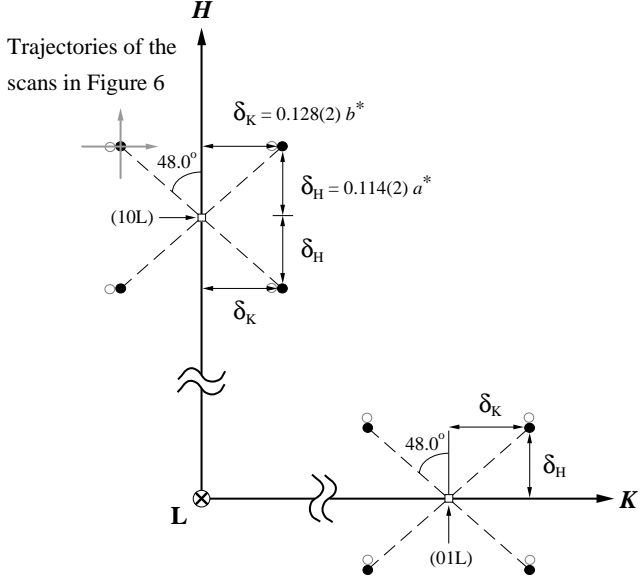}}
\vskip 5mm
\caption{
Positions of the {\it elastic} incommensurate spin density wave peaks
(closed circles) in a single structural twin domain, deduced from our
high resolution measurements.  $H$ and $K$ are in orthorhombic units.
The open circles represent SDW peaks originating from another twin
domain, which we measure simultaneously in our scans.  The
trajectories for the scans in Figures 6(A) and 6(B) are denoted by the
arrows through the position (1+$\delta_H$, -$\delta_K$, 0).
}
\label{Figure-elastic-positions}
\end{figure}

\noindent 
($100$) position, and not the $(010)$ position as shown in the upper
portion of Figure~5.  Therefore, each set of four peaks almost
certainly originate from a single structural twin domain.  In fact, we
also find incommensurate peaks about the $(01L)$ position (as shown in
the lower portion of Figure~5), but at odd-integer $L$ values, which
we discuss further below.  The intensities of the four peaks are
observed to be identical within the errors.  Surprisingly, we see that
the incommensurate wavevector is not precisely aligned with the
Cu-O-Cu tetragonal direction, but has different incommensurabilities
along the orthorhombic $H$ and $K$ directions:
$\delta_H\simeq0.114(2)~a^*$ and $\delta_K\simeq0.128(2)~b^*$.  This
corresponds to a rotation of the incommensurate wavevector by
3.3$^\circ$ with respect to the Cu-O-Cu direction.

We show in Figures 6(A) and 6(B) elastic scans through the
incommensurate elastic peak depicted by the trajectories illustrated
by the arrows in the upper left corner of Figure~5.  The measurements
have been performed with 5~meV neutrons and the horizontal
collimations are
\mbox{$30^\prime-20^\prime-S-20^\prime-B$}, where $S$ denotes the
sample and $B$ is a blank collimator.  The observed peaks are
extremely sharp and indeed are close to being resolution limited.  The
solid lines in the figure are 2D Gaussian line-shapes (with a delta
function in energy) convolved with the instrumental resolution.  The
double peak structure in Figure 6(B) results from structural
\linebreak

\begin{figure}
\vskip -5mm
\centerline{\epsfxsize=2.9in\epsfbox{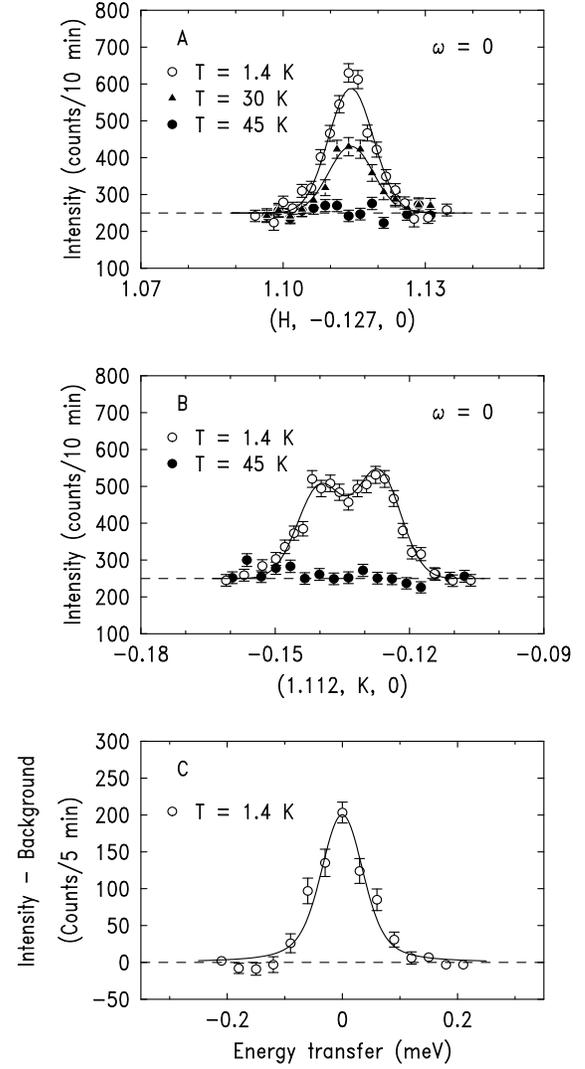}}
\vskip 5mm
\caption{
Elastic scans through the incommensurate peak position.  A) Scans
along the orthorhombic $H$ direction, as depicted in Figure~5, for
various temperatures.  B) Scans along the orthorhombic $K$ direction,
as depicted in Figure~5.  C) Scan of the energy transfer at a fixed
incommensurate {\bf Q}-position.  The solid lines denote the results
of fits to the cross section convolved with the resolution as
described in the text.
}
\label{Figure-elastic-scans}
\end{figure}

\noindent
twinning.  Fits to either Gaussian or Lorentzian intrinsic line-shapes
can adequately describe the data to within the errors and both
indicate that the in-plane static magnetic order is isotropically
correlated over distances greater than 400~\AA.  This lower bound on
the extent of the long-range in-plane spin correlations is limited by
the uncertainty in the instrumental resolution, which we could
quantify to within 10\% for the transverse direction.  Figure 6(C)
displays data in which the energy transfer is scanned while the
reciprocal space position is fixed at the peak position of the elastic
signal.  The observed energy width of the peak is also very narrow.
The solid line is
\linebreak

\begin{figure}
\vskip -3mm
\centerline{\epsfxsize=2.8in\epsfbox{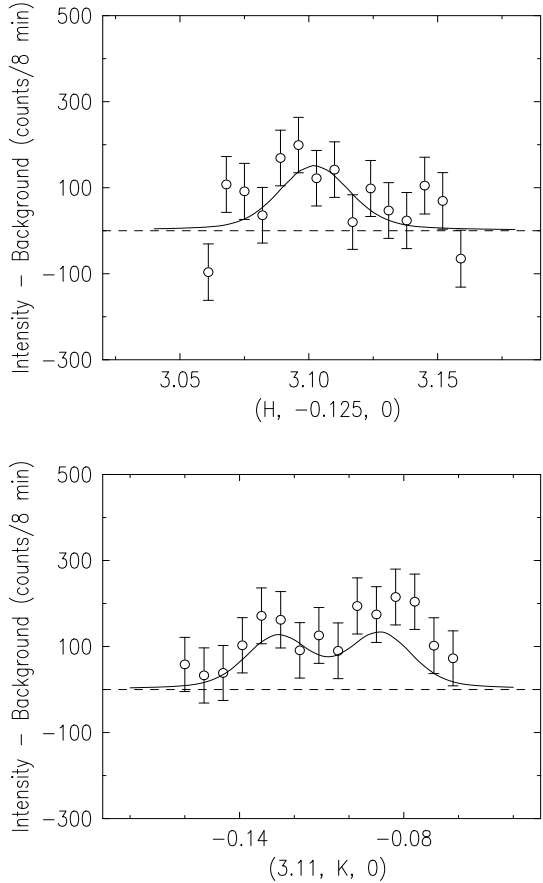}}
\vskip 4mm
\caption{
Elastic scans through the incommensurate peaks positions near
the $(300)$ antiferromagnetic Brillioun zone center position, measured
with 13.7~meV neutrons and horizontal collimations of
\mbox{$30^\prime-30^\prime-S-30^\prime-B$}.  The background signal at
$T=50$~K has been subtracted from the data taken at $T=2$~K.  The
solid lines denote the results a fit to the cross section convolved
with the resolution as described in the text.  The reduction in
intensity here relative to that measured around the $(100)$ position
is consistent with the Cu$^{2+}$ magnetic form factor in \LaPurens.  }
\label{Figure-300}
\end{figure}

\noindent 
a Lorentzian lineshape in energy transfer convolved with the
instrumental resolution, assuming the above narrow $q$-widths.  The
results of the fit indicate that the intrinsic energy width of the
peak is less than $\sim0.025$~meV HWHM at $T=1.4$~K; the instrumental
energy resolution is 0.075~meV HWHM.  We also observe incommensurate
peaks around the (300) position, albeit with much weaker intensity as
shown in Figure~7.  After taking into account the instrumental
resolution including the Lorentz factor and the crystal mosaic, we
find that the intrinsic signal around $(300)$ is smaller by a factor
of approximately 2.5 than that around $(100)$.  This is consistent
with the decrease in signal that is expected from the Cu$^{2+}$
magnetic form factor in undoped \LaPurens~\cite{Fretloft}.  By
contrast, for nuclear scattering the intrinsic signal should increase
by $Q^2\sim9$ between $(100)$ and $(300)$.  Thus, we conclude that the
observed incommensurate elastic scattering is magnetic in origin.

We next address the question of how the static spin arrangement is
correlated between successive copper oxide planes.  For this
experiment, the sample is remounted on the spectrometer so that the
$c$-axis and an orthorhombic in-plane axis are both in the scattering
plane.  In this way, reciprocal space positions corresponding to
$(H0L)$ may be reached, and, as a result of the twinning, $(0KL)$
positions are simultaneously measured.  In order to scan through the
incommensurate position, we must tilt the sample by 8$^\circ$ to bring
one of the incommensurate peaks into the scattering plane.  From
measuring the intensity of a ``$(10L)$-centered'' incommensurate peak
at $L=0$, we find that the intensity is a factor of 4 times smaller
than that measured in the $(HK0)$ configuration.  This is not
surprising, since the neutron spectrometer has a broad vertical
resolution which integrates the signal over a broad width of
reciprocal space perpendicular to the scattering plane; hence the
reduction in the detected intensity in this configuration suggests
that the intrinsic signal is composed of diffuse peaks along the
$L$ direction.  Indeed, in pure \LaPure the magnetic exchange between
nearest neighbor planes is four orders of magnitude smaller than the
exchange within the plane, so the spins exhibit observable
correlations along this out-of-plane direction only in the N\'{e}el
phase.

By fixing the in-plane component of the {\it \bf Q}-vector to lie on
the incommensurate position $(1-\delta_H,~\delta_K, L)$, we perform
scans along the $L$ direction through the peak.  Scans have been made
at $T=2$~K to measure the elastic magnetic intensity and have been
repeated at $T=50$~K to measure the background.  The background is not
strictly flat, so we performed $H$-scans at several fixed
$L$-positions and verified that the peaks are resolution limited for
the in-plane direction and that the appearance of this resolution
limited SDW peak is the only change in the scattering between 50~K and
2~K.  In Figure~8(A), we show the $L$-dependence of this scattering.
There is a clear modulation of the signal with maxima at even-integer
$L$ values.  The peaks are broad, suggesting short-range spin
correlations between the copper-oxide planes.  We have also performed
measurements with the alignment such
that the measured incommensurate
position was centered about the $(01L)$ position; that is, we fix the
{\it \bf Q}-vector to lie on the $(\delta_H, 1 -\delta_K, L)$
incommensurate position.  The $L$-dependence of the magnetic intensity
of this ``$(01L)$-centered'' domain is shown in Figure~8(B).

The intensity modulation of both the $(10L)$- and $(01L)$-centered
scattering is reminiscent of the ordered spin structure of the undoped
insulating parent compound \LaPurens, which has the spin direction
along $(010)$ and the antiferromagnetic propagation vector along
$(100)$.~\cite{Vaknin} Specifically, since \LaPure has an orthorhombic
structure, the magnetic exchange between planes is not perfectly
frustrated.  Thus, a particular 3D stacking arrangement of the spins
on nearest neighbor planes is observed, in which the inter-plane
nearest neighbor
\linebreak

\begin{figure}
\vskip -5mm
\centerline{\epsfxsize=3.1in\epsfbox{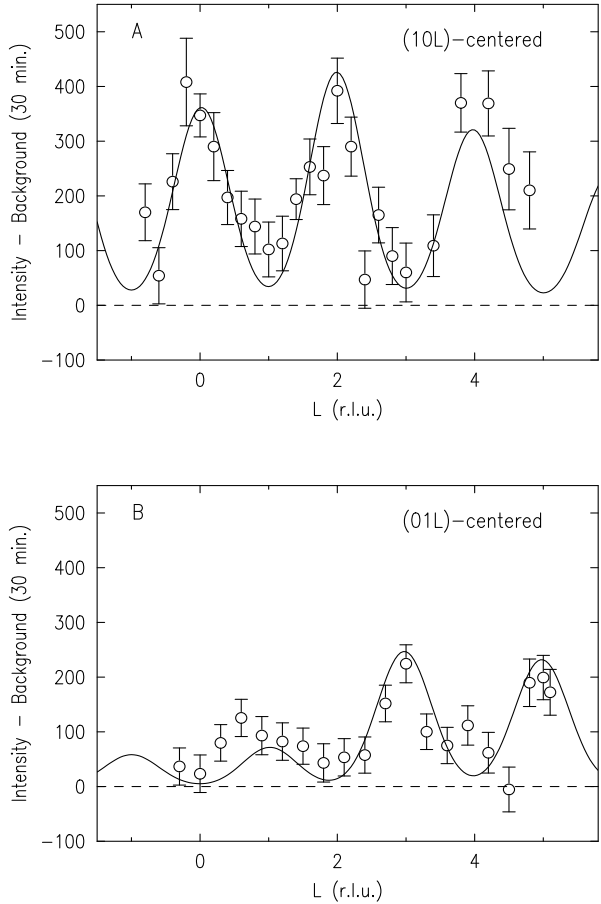}}
\vskip 5mm
\caption{
Elastic scans along the $L$ direction perpendicular to the CuO$_2$
planes, measured with 5~meV neutrons and horizontal collimations of
\mbox{$30^\prime-40^\prime-S-20^\prime-B$}.  A) Scan through the
incommensurate position centered about $(10L)$, scanning along
$(1-\delta_H,~\delta_K,~L)$.  B) Scan through the incommensurate
position centered about $(01L)$, scanning along
$(\delta_H,~1-\delta_K,~L)$.  The solid lines denote the results of a
fit assuming a model for the stacking arrangement and spin direction
identical to that in pure \LaPurens, as described in the text.
}
\label{Figure-L-dependence}
\end{figure}

\noindent
spins are aligned anti-parallel, yielding magnetic Bragg peaks at
$(H0L)$, with $H$ odd and $L$ even, and $(0KL)$, with $K$ odd and $L$
odd. In addition, the work of Thio \etns~\cite{Thio} has shown that in
pure \LaPure the Dzyaloshinski-Moriya anti-symmetric exchange term in
the spin Hamiltonian in the orthorhombic phase favors spin alignment
along the $K$ direction, which is also observed here.  The solid lines
in both panels of Figure~8 represent the results of fits to Gaussian
lineshapes in the $L$ direction (along with Bragg peaks for the
in-plane cross section as discussed above) convolved with the
instrumental resolution assuming a model for the stacking arrangement
and spin direction identical to that in pure \LaPure.  Here, the only
free parameters are the width of the peaks in $L$ and a single overall
intensity scale factor.  The agreement is clearly satisfactory, and
the fitted width indicates that, within a finite-size domain model,
the spins are correlated over $\sim13$~\AA~in the $L$ direction, or
across 2-3 CuO$_2$ planes.  As can be seen from Eq.~3, if the spins
order collinearly, the neutron cross section depends on the angle
$\theta$ between the {\bf Q} scattering vector and the ordered spin
direction by a factor which can be written as sin$^2$($\theta$).  The
relative intensities of all of the peaks in Figure 8, especially the
growth in the peak intensity for the $(01L)$-centered scattering for
increasing $L$ shown in Figure~8(B), are consistent with this
geometrical factor if the copper spin direction is fixed along the
$(010)$ direction.  We have explicitly verified via $K$-scans that the
$(01L)$-centered scattering at $L=3$ is significantly greater than
that at $L=1$ and is resolution limited for this in-plane direction.
Here, the magnetic form factor is assumed not to vary significantly
over the range of $L$ values scanned here, similar to the situation in
pure \LaPurens.~\cite{Fretloft,Shamoto} However, we do account for
instrumental resolution effects which weakly modify the intensity over
the course of the scan.  We conclude that the stacking arrangement of
the magnetically ordered planes in our \La sample is the same as that
of the undoped Mott insulator \LaPurens, even though the material is
superconducting and the magnetic scattering is incommensurate.  This
is direct and compelling evidence that the magnetism of the doped
superconductors is related to the magnetism in the undoped insulators
in a very specific way.

We show in Figure~1(B) the peak intensity of the elastic signal as a
function of temperature measured using both 5~meV and 13.7~meV
neutrons.  The fact that one obtains identical results for the
temperature dependences of the intensities with these two different
neutron energies and, concomitantly, energy resolutions implies that
the scattering is truly elastic.  We note that since the peak widths
do not vary with temperature to within the errors, the peak intensity
is proportional to the integrated intensity.  The intensity of the
elastic scattering turns on at approximately the same temperature as
the onset of superconductivity.  With decreasing temperature, the
intensity increases at first approximately linearly and then crosses
over to a saturation level at the lowest temperatures.  Noting that
the intensity of the magnetic scattering is proportional to the square
of the magnetic order parameter, we have plotted the square of the
Bardeen-Cooper-Schrieffer (BCS) weak coupling order parameter curve
together with the data using a \tm of 40.5~K equal to
\tcns(midpoint)$\simeq40.5$~K.  The good agreement demonstrates, at
the minimum, that the magnetism exhibits mean field behavior.  This is
very surprising given the two dimensionality of the ordered magnetism.

The size of the average ordered moment at base temperature has been
estimated several ways.  We compare the magnetic signal to that from a
crystal of pure \LaPurens, for which we know that the ordered moment
in the Neel state is $\sim0.55\mu_B$, on the same spectrometer under
identical conditions.  Also, we have made comparisons with a vanadium
standard and with the weak (004) Bragg reflection in the same crystal
for which extinction is not a significant factor.  All three
comparisons yield consistent results, with the size of the average
ordered moment of the SDW equal to $0.15\pm0.05\mu_B$ at $T=1.4$~K.
For the purposes of this estimate, we have assumed a simple model for
the SDW order, in the form of magnetic stripes with a width of three
spins, separated by non-magnetic anti-phase domain
walls.~\cite{Tranquada-Nd} In this case, one requires the presence of
two magnetic twin domains of stripes which are oriented at 90$^\circ$
relative to each other in order to account for the four incommensurate
peaks.  Even if we use a different model for the spin order, such as a
two-dimensional square grid of anti-phase domain walls, the variation
in the size of the calculated ordered moment is within the error as
quoted above.  We note that in both such simple candidate models, the
predicted third harmonic peaks are extremely weak
($\stackrel{_<}{_\sim} 5 \%$), and in fact we are not able to observe
them above the background.

\section{Discussion}

The existence of either short or long-range ordered incommensurate
spin density wave peaks now appears to be a common feature of the
\LaPurens-family of superconductors.  Such peaks have been observed
for several different Sr$^{2+}$ doping levels in Nd$^{3+}$ co-doped
samples of La$_{1.6-x}$Nd$_{0.4}$Sr$_x$CuO$_4$~\cite{Tranquada-Nd}, as
well as in purely Sr$^{2+}$ doped samples of \LaSrns~\cite{Kimura}.
Here we have shown that the SDW peaks are clearly seen in oxygen doped
samples, as well.  For the La$_{1.6-x}$Nd$_{0.4}$Sr$_x$CuO$_4$
samples, the SDW ordering and superconductivity coexist but have
different transition temperatures with \tm $>$ \tc for $x=0.12$ and
$x=0.15$, and \tm $\simeq$ \tc for $x=0.20$.  For \LaSrns, recent
experiments suggest that \tc $\stackrel{_>}{_\sim}$~\tm for
$0.10~{\Large
\stackrel{_<}{_\sim}}~x~{\Large \stackrel{_<}{_\sim}}~0.135$ with \tc
$\simeq$ \tm at $x=0.12$.  For both of these series of compounds,
which involve Sr$^{2+}$ doping, the quenched randomness of the
Sr$^{2+}$ dopants implies that the hole doping must be fairly
homogenous.  That is, macroscopic phase separation is prevented by the
tendency of the holes to remain in the vicinity of their donor atoms.
Hence, it is unlikely that that the magnetic order and the
superconductivity occur in macroscopically distinct phase-separated
regions of the single crystal samples. In fact, recent $\mu$SR and
magnetization experiments confirm that the previously studied
La$_{1.45}$Nd$_{0.4}$Sr$_{0.15}$CuO$_4$ crystal is a bulk
superconductor.~\cite{Tranquada-Nd,Nachumi-usr,Ostenson-mag} The
coexistence of SDW order and superconductivity for the many
superconducting compositions mentioned above argues strongly against
their resulting from sample inhomogeneity.  Thus, it appears that
there is true phase coexistence of superconductivity and long-range
incommensurate magnetic order in our La$_2$CuO$_{4.12}$ sample.  Of
course, we cannot rule out phase separation on short length scales,
such as into alternating superconducting and SDW lamellae.

The character of the SDW in the \La system is clearly of the same
fundamental nature as that of its Sr$^{2+}$-doped brethren, yet is
more ideal in several respects.  The superconducting \tc and SDW
ordering $T_m$ occur at the same onset temperature $\sim42~K$ to
within the uncertainties, and both temperatures are higher than the
highest respective temperatures for Sr$^{2+}$ doped materials.  It had
initially been believed that the SDW phase in
La$_{1.6-x}$Nd$_{0.4}$Sr$_x$CuO$_4$ ordered near $\sim50~K$; however,
a subsequent analysis shows that the magnetic order above 30~K is of
short-range, and the ordering below 30~K is more consistent with
glassy behavior.~\cite{Tranquada-glass} As discussed above for our
crystal of \Lans, we see very similar behavior for incident neutron
energies of 5~meV and 13.7~meV, which integrate over significantly
different energy windows (0.075 versus 0.45~meV HWHM, respectively,
with our particular choice for spectrometer collimations).  Thus, the
magnetic intensity measured throughout the entire temperature range
below $\sim40$~K is predominately static; $\mu$SR studies on our
samples would determine this more precisely.  It is apparent that both
\tc and \tm can be depressed in a particular system by, for example,
structural distortions, impurities, or dopant disorder.  Although we
cannot rule out a coincidence, the ordering of the spin density wave
at the same temperature as the superconductivity in our
\La sample and the observation that both have the highest ordering
temperatures for the respective phenomena in \LaPurens-based
materials suggest that the magnetism and the superconductivity are
synergistic; further, both are quite sensitive to disorder in the
CuO$_2$ planes.

It is now clearly demonstrated that the SDW exists in an orthorhombic
crystal structure.  In fact, the anisotropy between the in-plane $a$
and $b$ axes is reflected in the ordering pattern of the spins, which
have different incommensurabilities $\delta_H$ and $\delta_K$.  Thus,
as suggested by the previous work of Kimura \etns~\cite{Kimura} in the
\LaSr system, SDW order in the cuprates is a general phenomenon. The
magnitude of the long-range ordered moment at $T=1.4$~K for
superconducting \La is about 25\% of that in the undoped
antiferromagnet \LaPurens.  This is larger than the moment found in
the LTT phase of
La$_{1.48}$Nd$_{0.4}$Sr$_{0.12}$CuO$_4$~\cite{Tranquada-Nd};
therefore, the LTT structure does not enhance the magnetic ordering
relative to an orthorhombic structure.  Our \La sample is to-date the
only material in which the magnetic scattering is observed clearly to
saturate at low temperature, indicating that the long-range moment is
fully ordered.  Unlike the situation in the
La$_{1.6-x}$Nd$_{0.4}$Sr$_x$CuO$_4$ material we cannot yet draw any
conclusion regarding possible charge ordering.  Scans through the
positions of the charge order peaks which were observed in
La$_{1.48}$Nd$_{0.4}$Sr$_{0.12}$CuO$_4$~\cite{Tranquada-Nd,VonZimm}
exhibited a large sloping background in our \La sample due to the
close proximity of the scattering from the ordered distribution of
oxygen interstitials~\cite{Lee-Tewary}, precluding a firm
determination of the existence of charge ordering.  We note that since
neutrons scatter from small nuclear displacements induced by the
modulated charge density, the strength of the nuclear superlattice
peaks associated with any charge ordering may differ between the LTT
and LTO structures, even though the SDW scattering is similar.
Preliminary $^{63}$Cu NMR results on a small piece of our crystal by
Imai and coworkers~\cite{Imai} support an interpretation of the
appearance of charge ordering in La$_2$CuO$_{4.12}$ below $\sim55$~K
with a temperature dependence similar to that observed in
La$_{1.6-x}$Nd$_{0.4}$Sr$_x$CuO$_4$ by the same method~\cite{Hunt}.

The observation of static long range order is interesting in light of
the possibility of the existence of a quantum critical point in the
cuprate phase diagram.  Of course, since true long range order cannot
exist at non-zero temperatures in a 2D system with a continuous
symmetry, some spin anisotropy or 3D coupling must be present.  In
this case, it appears to be the Dzyaloshinskii-Moriya (DM) coupling
which breaks the continuous symmetry and makes possible a transition
to two-dimensional long-range order at non-zero temperature.  However,
an algebraic decay state is not ruled out, so long as the spins are
correlated over very large distances ($>400$~\AA).  We note that in a
model, such as the aforementioned stripe model, with antiphase domain
boundaries between ordered spins, the out-of-plane spin canting due to
the DM coupling will point in opposite directions on either side of a
domain wall.  This is a special case of the more general result that
due to the DM coupling, an antiferromagnetic SDW with
incommensurability $\vec{\delta}$ will give rise to a canted moment
with wavevector $-\vec{\delta}$.  As a consequence, each CuO$_2$ plane
will have no net ferromagnetic moment, consistent with preliminary
magnetization measurements~\cite{Thomas}, and in contrast to the weak
ferromagnetic moment observed in pure
\LaPurens~\cite{Thio}.  For the low-energy inelastic incommensurate
scattering, which is believed to be primarily two-dimensional, the
2~meV dynamical correlation length above \tm is exceedingly large,
$\stackrel{_>}{_\sim}125$~\AA, which is similar to that measured near
the special $x=1/8$ doping level in
\LaSrns~\cite{Yamada-doping}.  Recently, Aeppli and
co-workers~\cite{Aeppli-singular} have reported pre-divergent behavior
in the dynamical properties of La$_{1.86}$Sr$_{0.14}$CuO$_4$, albeit
with a shorter correlation length of 30~\AA.  In this work we clearly
see an ordered magnetic phase at an intermediate doping level of the
cuprate phase diagram.  Even though the spins are correlated
three-dimensionally, the fact that the in-plane correlations have
long-range order and the between-plane correlations do not,
necessitates that the phase transition is driven by the 2D
correlations of a single CuO$_2$ plane.  Clearly, studies of the
associated 2D critical fluctuations would be most interesting.

The absence of extinction of the 2 meV spin excitations as one cools
below \tc may now be simply explained.  Since our \La sample has a
long-range ordered ground state, spin wave excitations, which have a
temperature independent generalized susceptibility at low $T$, are
expected once the sample is cooled below \tmns.  This is clearly
consistent with the data in Fig.~4(B).  In addition, Kimura
\etns~\cite{Kimura} have shown that \LaSrns with $x=0.15$ does not
support a static spin density wave.  Again, this is consistent with
the inelastic scattering data~\cite{Yamada-gap} which show a complete
gap in the spin excitations at low temperatures for energies $\leq
3.5$~meV.  Our measurements also highlight the need to carry out high
resolution measurements of the inelastic magnetic scattering in
La$_2$CuO$_{4.12}$ to search for dynamic manifestations of the
anisotropic incommensurabilities $\delta_H$ and $\delta_K$ seen in the
elastic scattering.

The observed characteristics of the SDW argue against an itinerant
electron description of the incommensurate magnetism.  Specifically,
it is difficult to see how a first principles' delocalized model would
predict both interplanar spin correlations and a spin direction which
are identical to those in the undoped insulator, \LaPurens.  Also,
assuming that the elastic magnetic peaks originate from particle-hole
scattering across the Fermi surface of a superconducting condensate,
the wavevector should span from node to node of the presumed
$d_{x^2-y^2}$ superconducting order parameter.  Such a wavevector
would lie along the $(H00)$ direction; we find no observable
scattering along this direction, but only at the aforementioned
incommensurate positions.  Therefore, even in the superconductor, the
magnetism reflects in detail its Mott-insulator parentage.
Additionally, our observations are incompatible with any spiral-type
incommensurate modulation of the spin direction.~\cite{Siggia}

A plausible explanation of the SDW peaks is a stripe model similar to
that proposed to describe
La$_{1.6-x}$Nd$_{0.4}$Sr$_x$CuO$_4$~\cite{Emery,Tranquada-Nd} and as
simulated by Kim \etns~\cite{Kim}.  If stripes are indeed present,
then our data impose several important constraints on the model.
First, the stripes are not perfectly aligned with the Cu-O-Cu
direction, but are slightly rotated.  That is, they are tilted by
$\sim3^\circ$ from perfect orientational alignment with the underlying
spin lattice.  This is equivalent to a kink by 1 lattice constant
every 17 unit cells.  Second, in order to give the observed
$L$-dependence, the stripes on neighboring planes must be parallel
with each other with the charge domain walls projected nearly on top
of each other.  This preserves the phase of the 3D spin propagation
wavevector.  Third, there must exist two types of magnetic twin
domains composed of stripes in orthogonal directions with
approximately equal populations.  Alternatively, a grid model, in
which there is a rectangular array of antiphase domain walls running
along the orthorhombic directions, is also consistent with the data.
Such a model would more naturally account for the differing $\delta_H$
and $\delta_K$ incommensurabilities by having different domain wall
spacings along the two orthorhombic directions.  Further experimental
investigation of any associated charge ordering is crucial in order to
distinguish between these and possibly other models.

\section{Acknowledgments}

We gratefully acknowledge G.~Aeppli, F.C.~Chou, R.J.~Christianson,
V.J.~Emery, H.~Fukuyama, M.~Greven, K.~Hirota, T.~Imai, S.A.~Kivelson,
P.A.~Lee, P.K.~Mang, A.~Tewary, J.M.~Tranquada, and B.O.~Wells for
valuable discussions.  The present work was supported by the US-Japan
Cooperative Research Program on Neutron Scattering.  The work at MIT
was supported by the NSF under Grant No. DMR97-04532 and by the MRSEC
Program of the National Science Foundation under Award No.
DMR98-08941.  The work at Tohoku has been supported by a Grant-in-Aid
for Scientific Research of Monbusho and the Core Research for
Evolutional Science and Technology (CREST) Project sponsered by the
Japan Science and Technology Corporation.  The work at Brookhaven
National Laboratory was carried out under Contract No.
DE-AC02-98CH10886, Division of Material Science, U.S.  Department of
Energy.  The work at SPINS is based upon activities supported by the
National Science Foundation under Agreement No.  DMR-9423101.



\end{document}